# Half-Metallic Silicene and Germanene Nanoribbons: towards High-Performance Spintronics Device

Yangyang Wang,[1] Jiaxin Zheng,[1,2] Zeyuan Ni,[1] Ruixiang Fei,[1] Qihang Liu,[1] Ruge Quhe,[1,2] Chengyong Xu,[1] Jing Zhou,[1] Zhengxiang Gao,[1] and Jing Lu[1,*]

[1] State Key Laboratory for Mesoscopic Physics and Department of Physics, Peking University, Beijing 100871, P. R. China

[2] Academy for Advanced Interdisciplinary Studies, Peking University, Beijing 100871, P. R. China

*Corresponding author: jinglu@pku.edu.cn

*Abstract:* By using first-principles calculations, we predict that an in-plane homogenous electrical field can induce half-metallicity in hydrogen-terminated zigzag silicene and germanene nanoribbons (ZSiNRs and ZGeNRs). A dual-gated finite ZSiNR device reveals a nearly perfect spin-filter efficiency of up to 99% while a quadruple-gated finite ZSiNR device serves as an effective spin field effect transistor (FET) with an on/off current ratio of over 100 from *ab initio* quantum transport simulation. This discovery opens up novel prospect of silicene and germanene in spintronics.

## 1. Introduction

Because of the extremely high carrier mobility and long spin relaxation time and length, graphene could play a vital role in nanoelectronics and nanospintronics. High-performance radio frequency graphene transistor has been fabricated by using the extremely high carrier mobility.[1-3] An in-plane transverse electrical filed can cause half-metallicity in hydrogen-terminated zigzag graphene nanoribbons (ZGNRs)[4,5] due to the existence of unique edge magnetic state according to theoretical calculations, suggesting potential of ZGNRs in spintronics. Giant magnetoresistance (MR) is also predicted for magnetic ZGNRs.[6-8] The experimental confirmation of these exciting predictions suffers from the fabrication difficulty of GNRs with a smooth edge. With the development of experimental technique, well-defined ZGNRs have been experimentally fabricated very recently.[9,10] Especially, the observed double Raman G-peak is in good agreement with the calculation for ZGNRs.[11]

Silicon and germane are the two most important semiconducting materials. The boom of graphene research stimulates effort to search for its two-dimensional honeycomb single-atom layer analogue in Group IV, such as, silicene and germanene. Synthesis of Mg-doped,[12] and hydrogenated[13] silicene, and pristine silicene nanoribbons[14-18] have been reported. Both silicene and germanene have a Dirac cone in the electronic structure, and extremely high carrier mobility is expected (similar to graphene).[16, 19-21] Silicene-based field effect transistors (FET) have been proposed, with an electrical field-induced tunable band gap.[21] Silicene is probably more easily fitted into current Si-based semiconductor devices compared with graphene.[14] In addition, silicon has a longer spin-diffusion time ($\tau_s$ = 1 ns at 85 K[22] and 500 ns at 60 K[23]) and spin coherence length ($l_s$ = 10,[22] 350,[23] 2000 μm[24])[25] compared with graphene ($\tau_s$ = 0.1 ns and $l_s$ = 1.5 and 2 μm at room temperature[26]) and appears more suitable for spintronics application than graphene. Zigzag silicene and germanene nanoribbons (ZSiNRs and ZGeNRs) are predicted to have an antiferromagnetic (AFM) ground state as a result of edge magnetic state coupling.[20, 27]

In this article, we provide an investigation on the effects of in-plane transverse external electric field ($E_{ext}$) on hydrogen-terminated ZSiNRs and ZGeNRs by using spin-polarized density functional theory (DFT) and non-equilibrium Green's function (NEGF) methods. Interestingly, electrical field-induced half-metallicity is revealed in both ZSiNRs and ZGeNRs, and the critical electrical field to induce half-metallicity decreases with the increasing ribbon width. Subsequently, we design a model, where a finite ZSiNR is connected to two homogeneous electrodes to device application. *Ab initio* quantum transport simulation shows that when a pair of gate electrodes is applied on the two sides of the ZSiNR, nearly perfect polarized current is available. Inspired by the experiment where two pairs of side gates were fabricated for graphene strip,[28] we then apply two pairs of gate electrodes on the two sides of the ZSiNR, and find that this device operates as an effective spin FET. These findings make ZSiNRs and ZGeNRs promising candidates for both electronic and spintronic devices.

## 2. Computational methods

Geometry optimization and electronic structure of the ZSiNRs and ZGeNRs are carried out by using all-electron numerical double atomic basis set plus polarization (DNP) implemented in the DMol$^3$ package[29, 30]. The criterion of maximum force during optimization is 0.1 eV/Å. A $1 \times 1 \times 50$ Monkhorst-Pack[31] k-points grid is used in the first Brillouin zone sampling. The generalized gradient approximation (GGA) of the Hamprecht-Cohen-Tozer-Handy (HCTH) form[32] to the exchange-correlation functional is adopted since it well reproduces the band gap (1.17 eV) of bulk silicon. To analyze the origination of half-metallicity, we calculate the orbitals of the conduction and valence bands by using the ultrasoft pseudopotential plane-wave basis set implemented in the CASTEP package.[33] The plane-wave cutoff energy is 180 eV and the form of GGA is Perdew–Burke–Ernzerhof (PBE) functional.[34] Transportation properties are simulated by using the DFT coupled with NEGF formalism implemented in the ATK 11.2 package[35-37] within the local spin density approximation (LSDA) to the exchange-correlation functional. A single-$\zeta$ plus polarization (SZP) orbital basis set is employed, and the Monkhorst-Pack[31] $1 \times 1 \times 100$ k-points grid is used to sample the one-dimensional Brillouin zone. The temperature is 300 K. Effects of gate are calculated by solving the Poisson equation self-consistently instead of simply lifting the central region's chemical potential.

## 3. Results and discussion

### 3.1 Electronic band structure under a transverse electrical field

We use *n*-ZSiNR (*n*-ZGeNR) denote a ZSiNR (ZGeNR) with *n* zigzag chains across the width. The fully relaxed structures of the hydrogen passivated ZSiNRs and ZGeNRs are puckered slightly with buckling distances of $h = 0.460$ and $0.676$ Å, respectively, which are in good agreement with the previous results.[20] The lattice constants of the ZSiNRs and ZGeNRs are $a = 3.866$ and $4.063$ Å, respectively. There is a doubly degenerate flat edge-state band at the Fermi level ($E_f$) (not shown) when spin is not considered, which results in a very large density of states at $E_f$. ZSiNRs and ZGeNRs will become magnetic by an infinitesimal on-site Coulomb repulsion. When spin freedom is included, ZSiNRs and ZGeNRs have a semiconducting ground state where the spin orientations at each zigzag edge are parallel but antiparallel between the two edges (antiferromagnetic (AFM)), as shown in Fig. 1(a). Another magnetic state with spin ferromagnetically (FM) coupled at the two edges is metallic and slightly higher in energy. We find that the magnetic coupling strength between the two edges is generally larger in the ZSiNR than that in the ZGeNR with the same zigzag chain number $n$. The AFM-FM energy differences per edge atom are $\Delta_{AFM-FM} = 4.7, 6.7, 6.0, 11.5, 48.4$ meV for the 4-, 5-, 6-, 8-, and 10-ZSiNR, which are apparently larger than those of the ZGeNRs (3.3, 4.3, 2.0, 1.3, 0.9 meV for the 4-, 5-, 6-, 8-, and 10-ZGeNR, respectively) and ZGNRs (3.8, 6.2, 5.6, 1.8, 1.1 meV for the 4-, 5-, 6-, 8-, and 10-ZGNR, respectively) with the same zigzag chain number. Therefore, the AFM $n$-ZSiNR is more stable than the AFM $n$-ZGeNR and $n$-ZGNR against temperature or external magnetic field perturbation. The AFM-FM energy difference of the $n$-ZSiNR starts to decrease with $n$ from $n = 12$. Notably, the AFM-FM energy difference per edge atom for the 10-ZSiNR (48.4 meV) surpasses the room temperature (26 meV), suggesting that the AFM state of the 10-ZSiNR can survive the room temperature. Interestingly, we find that $\Delta_{AFM-FM}$ firstly increases and then reduces with the increasing ribbon width in all calculations of ZGNRs, ZSiNRs and ZGeNRs. According to general understanding, the energy difference $\Delta_{AFM-FM}$ should reduce with the increasing ribbon width $n$.[38, 39] Lee $et\ al.$[38] demonstrate it by calculating $\Delta_{AFM-FM}$ of $n = 4, 8$, and 16, while Son $et\ al.$[39] calculate $\Delta_{AFM-FM}$ of $n= 8, 16$, and 32. As $n$ increases, there are two factors affecting $\Delta_{AFM-FM}$: one is the gradually diminishing constructive (destructive) interference between the AFM (FM) magnetic tails, and the other is the increasing absolute value of exchange energy for each state due to the increasing $n$. When the width is within the decay length of the spin-polarized edges state, the latter plays a leading role. Due to the destructive interference at the inner sites of nanoribbons for the FM ordering and enhanced magnetic moments at the inner sites of nanoribbons for the AFM ordering, the AFM gains larger exchange energy than the FM. Thus, $\Delta_{\text{AFM-FM}}$ becomes larger as $n$ increases when $n$ is small. However, when the width is larger than the decay length of the tails, the former factor will dominate. As $n$ increases, $\Delta_{\text{AFM-FM}}$ decreases and eventually vanishes.

The electronic structures of the AFM ZSiNRs and ZGeNRs are characterized by a direct band gap at $k = 0.75\pi/a$ and a degeneracy for two spins in all bands, as shown in the left panel of Fig. 1(b). These band gaps originate from the inequivalency of the two sublattices caused by the magnetic ordering (Fig. 1(a)). The size of the direct band gap is inversely proportional to the ribbon width for both kinds of nanoribbons since the strength of the staggered potentials decreases with the increasing ribbon width. The $n$-ZSiNR has a larger band gap than the $n$-ZGeNR, suggestive of a larger difference in the staggered potentials in the $n$-ZSiNR than in the $n$-ZGeNR.

In Fig. 1(c), we plot the zero-field $\alpha$-spin and $\beta$-spin orbitals of the valence and conduction band of the 6-ZSiNR. The orbitals are shown as the square of the absolute value of the wavefunction summed over all $k$-points. It is apparent that the oppositely oriented spin states of both the conduction and valence bands are localized at the opposite edges of the nanoribbon, and in the same edge the spin orientations in the conduction and valence bands are opposite. According to electrostatics, when a transverse external electric field is applied from the left to the right, the energies of $\alpha$-spin and $\beta$-spin states on the left edge of ZSiNRs or ZGeNRs drop ($e\Delta V < 0$), while their energies on the right edge rise ($e\Delta V > 0$). As a result, the $\beta$-spin states in the valence and conductive bands approach each other around $E_f$, whereas the $\alpha$-spin states are separated from each other. Consequently, ZSiNRs and ZGeNRs would become half-metallic when the electrical field is large enough.

Fig. 1(b) also shows the band structures of the 6-ZSiNR under $E_{\text{ext}} = 0.1$ and 0.25 V/Å. The spin degeneracy of the conduction and valence bands is indeed lifted by the transverse electrical field. As expected, the band gap of $\beta$-spin state decreases and finally closes under $E_{\text{ext}} = 0.25$ V/Å, while that of $\alpha$-spin state increases slightly relative to the zero-field value. The changes in the band gap of the two spin states with the electrical field for the checked ZSiNRs and ZGeNRs are displayed in Fig. 2. The band gaps of the $\beta$-spin state always decrease initially with the increasing $E_{\text{ext}}$ and finally close, whereas those of the $\alpha$-spin state initially increase and then decrease but still larger than their respective zero-field values even at $E_{ext}$ under which the band gaps of $β$-spin state vanish. Therefore, electrical field-induced half-metallicity is well established for both ZSiNRs and ZGeNRs. The minimum electrical field $E_{ext}^{Min}$ needed to achieve half-metallicity decreases with the increasing width of the ZSiNRs and ZGeNRs. $E_{ext}^{Min}$ is 0.5, 0.4, 0.25, 0.2, and 0.16 V/Å for the 4-, 5-, 6-, 8-, and 10-ZSiNR, respectively, and 0.4, 0.22, 0.2, 0.15, and 0.12 V/Å for the 4-, 5-, 6-, 8-, and 10-ZGeNR, respectively. $E_{ext}^{Min}$ for the *n*-ZSiNR is larger than that for the *n*-ZGeNR because the larger the band gap, the larger the required voltage drop between the two edges to shift the edge states into half-metallicity. Half-metallicity of the 6-ZSiNR can be maintained within the scope from 0.25 to 0.4 V/Å range and then the band gap of $β$-spin state is reopened (Fig. 2(b)). Eventually $α$- and $β$-spin states become degenerate at all bands with an indirect band gap and meanwhile magnetism is quenched from $E_{ext}$ = 0.5 V/Å, that is, the ground state has changed into the nonmagnetic state. According to the results of calculation by Yi Ding et al.,[27] 6-ZSiNR becomes half-metal when $E_{ext}^{Min}$ = 0.15 V/Å, relatively much smaller than our results, which is due to the different density functionals we adopted. The exchange-correlation functional PBE they used resulting in 6-ZSiNR with a direct band gap of about 0.17 eV while as for HCTH form we used the band gap of 6-ZSiNR is 0.44 eV. We test the different exchange-correlation functionals on calculation of bulk silicon, such as PBE form[34] in the SIESTA code[40] they adopted or HCTH form[41] in the DMol$^3$ package[29, 30], its calculated band gap for bulk silicon is 0.60 eV or 1.35eV, that is, the former functional dramatically undervalues the band gap (1.17 eV) while the latter well reproduces it. In addition, we reveal that the half-metallic behavior of ZSiNRs and ZGeNRs only maintains in a finite scale of the transverse electric field and then disappears when the electric field increased continually, which can provide an parameter to scale the robustness of ZSiNRs and ZGeNRs' half-metallicity.

**3.2 Spin filter**

We investigate the transport properties of a finite 4-ZSiNR in the presence of a transverse electrical field. For sake of computational convenience, nonmagnetic semi-infinite 4-ZSiNR is selected as metallic electrodes. The channel contains 21 unit cells of the ZSiNR and has a length of 81.2 Å. The distance between the two gates situated on both sides of the nanoribbon is $d_0 = 22$ Å, and the width of the nanoribbon is $d_i = 14$ Å. The transverse electric field generated by the gates and applied to the ZSiNR can be written as $E_{gate} = E_{ext} = \frac{V_g}{d_0}$.

First, we apply a pair of electrodes labeled as $-V_g/2$ and $V_g/2$, respectively, on the two sides of the nanoribbon (dual-gated), as illustrated in Fig. 3(a). We expect this device to operate as a spin-filter under a gate electrical field. Fig. 3(b) presents the spin-resolved transmission spectrum of this device under $E_{gate} = 0.3$ V/Å. There is a large peak for α-spin around $E_f$, and by sharp contrast a clear gap of about 0.17 eV appears around $E_f$ for β-spin, a result qualitatively consistent with the band calculation under this electrical field. The transmission coefficient at $E_f$ for α-spin $T(E_f)$ is much larger than that β-spin. We calculate the spin filter efficiency (SFE) at zero bias by using the formula as follows:

$$\text{SFE} = \frac{T_\beta(E_f) - T_\alpha(E_f)}{T_\alpha(E_f) + T_\beta(E_f)} \times 100\% \tag{1}$$

The resulting SFE is up to 99.2% under $E_{gate} = 0.3$ V/Å. The highly spin polarization is also reflected from the spin-resolved transmission eigenstate at $E_f$ and the Γ point in *k*-space, as shown in Fig. 3(c). The transmission eigenvalue of the *β*-spin is 1.943, in which case the scattering is very weak and most of the incoming wave is able to reach to the other lead. Consistent with the band calculation, this transmission channel is chiefly distributed along the two edges, On the contrary, the transmission eigenvalue of the *α*-spin is 0.008, and the corresponding incoming wave function is apparently scattered and unable to reach to the other lead.

The change of $T(E_f)$ and SFE as a function of $E_{gate}$ is shown in Fig. 3(d). Both the $T(E_f)$ and SFE are antisymmetric or symmetric about $E_{gate} = 0$ since *n*-ZGNR with even *n* is symmetric about its vertical midplane. The $T(E_f)$ between the two spins has a difference even at small $E_{gate} = -0.05$ or 0.05 V/Å, with a SFE = ±62.3%, and the difference becomes more and more significant with the increasing $E_{gate}$, a behavior consistent of the electrical field-induced change in the band gap of ZSiNRs. SFE is nearly saturated (±99%) from $|E_{gate}| > 0.2$ V/Å. Therefore the dual-gated finite ZSiNR can serve as a nearly perfect spin-filter, with sign switchable by altering the electric field direction.

### 3.3 Spin field effect transistor

We apply two pairs of electrodes on the two sides of the 4-ZSiNR (quadruple-gated), as illustrated in Fig. 4(a). We fix the electrical field of the left pair of electrodes ($E_{LG}$) and modulate the right one ($E_{RG}$). When $E_{RG} = E_{LG}$, the left and right parts of the nanoribbon allow the same spin to transport along the edges. This device degenerates into a spin-filter with $E_{gate} = E_{RG} = E_{LG}$. When the direction of the electrical field of the right pair of electrodes is reversed, the sign of the allowed travelling spin in the right part of the nanoribbon is reversed and contrary to that in the left part, resulting in a possible blockade of the transmission of both spins. As a result, the current of this device is expected to be forbidden in this case. Therefore, through altering $E_{RG}$ and thus altering the spin state, the quadruple-gated device can operate as a spin FET.

We fix the electrical field of the left pair of electrodes at 0.3 V/Å and modulate the right one. The total transmission spectrum under $E_{RG} = \pm 0.3$ V/Å is compared in Fig. 4(b). As expected, there is a large peak existing around $E_f$ when $E_{RG} = 0.3$ V/Å, but there is only small peak instead around $E_f$ when $E_{RG} = -0.3$ V/Å. Therefore, the total transmission coefficient at $E_f$ ($T(E_f)$) and conductance ($\sigma = (2e^2/h) \times T(E_f)$) under $E_{RG} = 0.3$ V/Å are apparently larger than those under $E_{RG} = -0.3$ V/Å, indicative of function of transistor. This difference in $T(E_f)$ is also reflected from the transmission eigenstate under $E_{RG} = \pm 0.3$ V/Å. The transmission eigenstate at $E_f$ and the Γ point under $E_{RG} = 0.3$ V/Å is equivalently shown in Fig. 3(c), while that under $E_{RG} = -0.3$ V/Å is plotted in Fig. 4(c). The conductive transmission channel is open for $\beta$-spin under $E_{RG} = 0.3$ V/Å, but both conductive transmission channels are basically closed nearly at the interface of the two reverse electrical fields under $E_{RG} = -0.3$ V/Å with similar transmission eigenvalues of about 0.16. Fig. 4(d) shows the value of $T(E_f)$ as a function of $E_{RG}$. As $E_{RG}$ varies from -0.3 to 0.3 V/Å, the total $T(E_f)$ increases generally. Therefore, through altering $E_{RG}$ and thus altering the spin state, the device shown in Fig. 4(a) operates as a spin FET. The maximal on/off conductance ratio of the present device is 18.

The low on/off ratio is ascribed to the quiet short channel (38.3 Å) controlled by each pair of electrodes in this simulation limited by the computational resource, which gives rise to a certain amount of leakage current on the off-state due to tunneling effect. If the channel length is increased, a higher on-off ratio is expected because the leakage of the two spin currents will both be reduced. When the channel length controlled by each pair of electrodes is increased to 76.6 Å, we can estimate that under zero bias the $T(E_f)$ of the off-state (twice $T_α(E_f)$ under $E_{gate}$ = 0.3 V/Å in Fig. 3(b)) will be reduced to about 0.016, the on/off ratio is therefore reasonably estimated to be 117.

Generally, the SFE of spin filter and on/off ratio of spin FET tend to be degraded with the increasing bias voltage.[8] We calculate the transport properties of the dual-gated devices under a fixed finite bias voltage of 0.05 V and $E_{gate}$ = 0.3 V/Å. The spin-resolved current $I_σ$ under a bias voltage $V_{bias}$ and a gate voltage $V_g$ is calculated with the Landauer-Büttiker formula[42]:

$$I_σ(V_{bias}, V_g) = \frac{e}{h} \int_{-∞}^{+∞} \{T_σ(E, V_{bias}, V_g)[f_L(E - μ_L) - f_R(E - μ_R)]\}dE, \quad (2)$$

Where $T_σ(E, V_{bias}, V_G)$ is the spin-resolved transmission probability at a given bias voltage $V_{bias}$ and gate voltage $V_g$, $f_{L/R}$ the Fermi-Dirac distribution function for the left (L)/right (R) electrode, $μ_L/μ_R$ the electrochemical potential of the L/R electrode, and $σ$ a spin index. When $E_{gate}$ = 0.3 V/Å, the $α$-spin and $β$-spin currents are 1.58 μA and 4.4 ×10$^{-2}$ μA, respectively. We calculate SFE at the finite bias voltage by using the formula as follows:

$$SFE = \frac{I_α - I_β}{I_α + I_β} × 100\% \quad (3)$$

The calculated SFE is slightly decreased to 94.6%, but remains very high. Through altering $E_{gate}$ from -0.3 to 0.3 V/Å, SFE can be modulated from -94.6% to 94.6%. The on/off ratio of the spin FET at a bias voltage of 0.05 V is expected to be slightly reduced compared with that under the zero bias voltage, but still high enough.

In our spin FET device composed of ZSiNRs, the current is modulated through applying a transverse electric field. The current in ZSiNRs also can be modulated by applying a magnetic field. Xu *et al.*[43] study the transport properties of finite ZSiNRs connecting two planar silicene electrodes and find that by using a magnetic field to switch the magnetic coupling between the two edges, a maximum optimistic MR up to 1960% is obtained because of a large current difference between the semiconducting AFM and metallic FM states. Besides, Kang *et al.*[44] predicted a MR of up to 10$^6$% in even-*n* ZSiNRs through switching the spin configuration of the two electrodes from parallel to anti-parallel configuration. The various measures to modulate the devices based on silicene show that it could provide flexibility for device design and is of great potential in the future nanoscale spintronics. It is notable that the choice of exchange-correlation functional affects the half-metallicity in ZGNRs. On the basis of B3LYP results, finite ZGNRs behave as half-semiconductors.[45] The smaller band gap is not closed completely, and no half metallicity is observed in ZGNRs. This phenomenon may remain in ZSiNRs and ZGeNRs when B3LYP functional is used. Even so, when a transverse electric field is applied, the highly spin-polarized current should appear due to the highly spin-dependent band gap although the SFE of the spin filter and the on/off ratio of the spin FET would be decreased.

## 4. Conclusions

We reveal that hydrogen-terminated zigzag silicene and germanene nanoribbons can become a half-metal by applying a transverse external electric field. Compared with their graphene counterparts, hydrogen-terminated zigzag silicene nanoribbons have a longer spin relaxation time and spin-diffusion length, higher magnetic stability, and probably better compatibility with the existing semiconductor technology. We simulate the transport properties of a dual-gated and quadruple-gated finite zigzag Si nanoribbon and find they can work as a perfect spin-filter and an effective spin transistor without magnetic contact, respectively. Therefore, a new path may be opened to explore spintronics based on silicene and germanene, especially in view of the fact that ZSiNR has been fabricated on Ag(110) surface.[14-18]

**Acknowledgement** This work was supported by the NSFC (Grant No. 10774003), National 973 Projects (No. 2007CB936200, MOST of China), Fundamental Research Funds for the Central Universities, National Foundation for Fostering Talents of Basic Science (No. J0630311), and Program for New Century Excellent Talents in University of MOE of China.

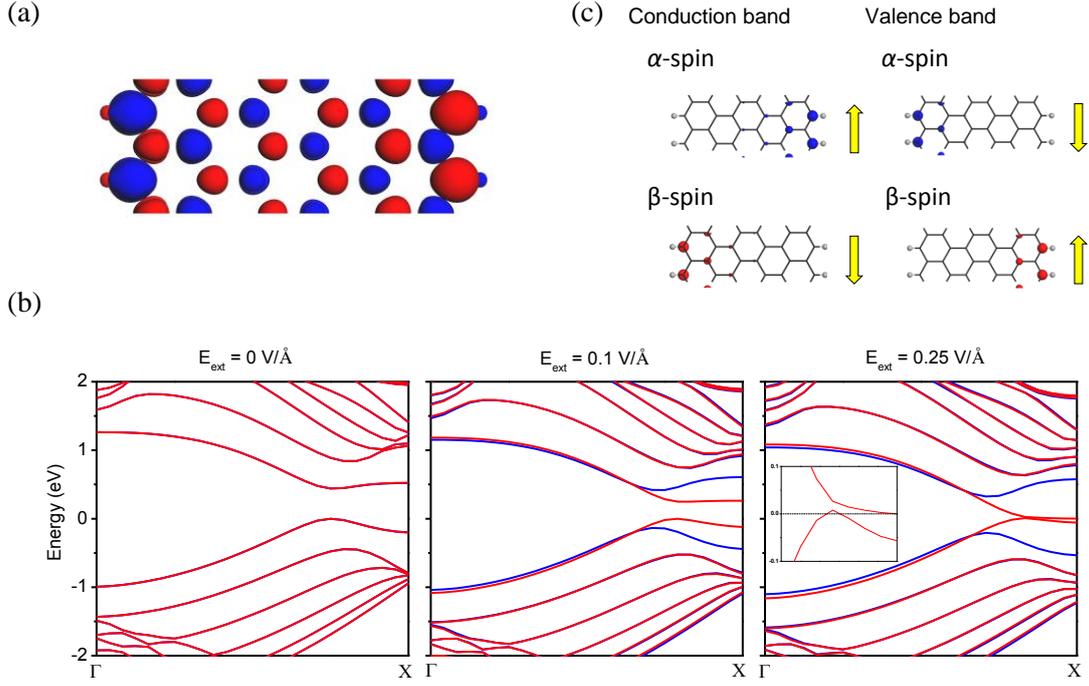

Fig. 1 Electronic properties of the 6-ZSiNR in the ground state. (a) Spatial spin density distribution. The isovalue is 0.006 a.u. (b) Spin-resolved band structures under $E_{ext}$ = 0, 0.1, and 0.25 V/Å, respectively. Inset: the band structure with $E_{ext}$ = 0.25 V/Å in the range of $|E| <$ 0.1 eV and $0.7\pi/a \leq k \leq \pi/a$ (the horizontal line is $E_f$). The valence top or $E_f$ is set to zero. (c) *α*-spin and *β*-spin orbitals of the conduction and valence band, shown as the square of the absolute value of the wavefunction summed over all *k*-points. The isovalue is 0.275 a.u. The yellow arrow represents the energy shift direction of the spin states under a transverse electrical filed. Blue and red denote *α*-spin and *β*-spin, respectively.

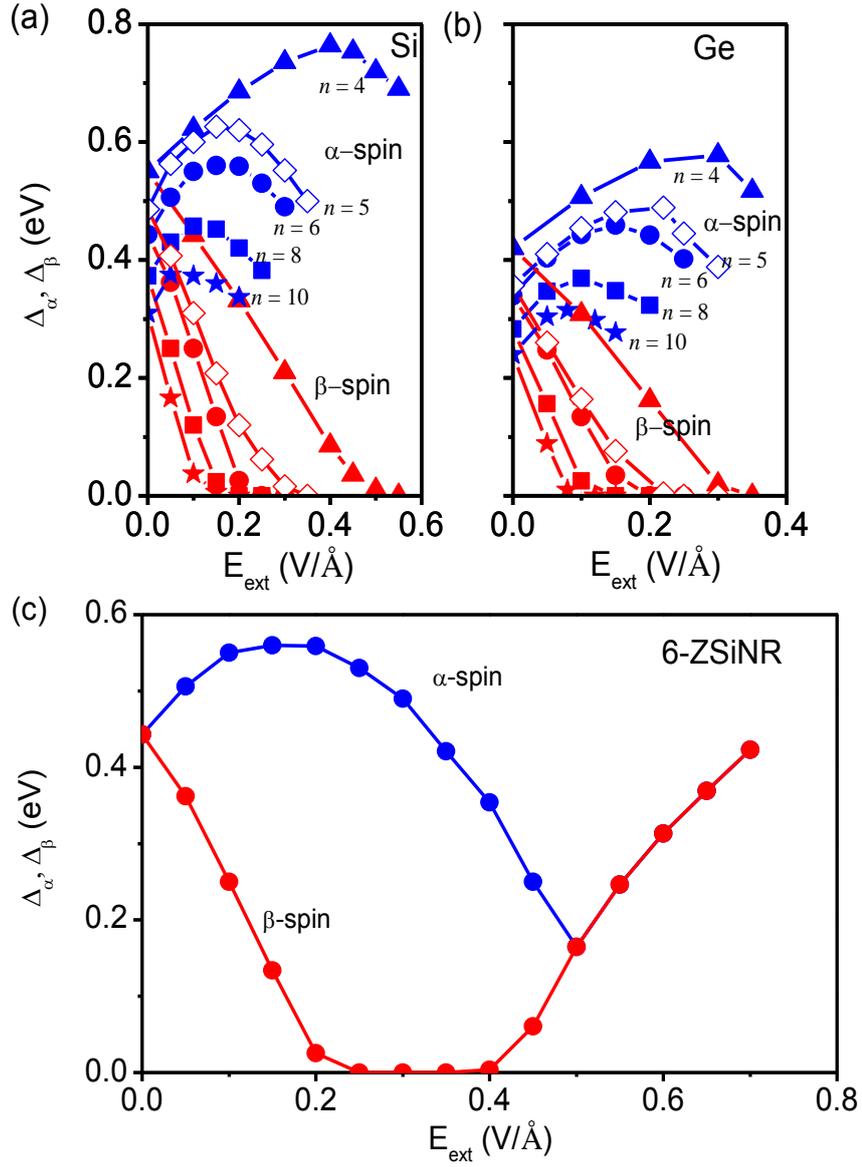

Fig. 2 (a) Spin-resolved band gaps for the $n$-ZSiNR with $n = 4, 5, 6, 8$, and $10$ as a function of $E_{ext}$. (b) Same as (a) but for the $n$-ZGeNR. (c) Spin-resolved band gaps for the 6-ZSiNR in a larger range of $E_{ext}$.

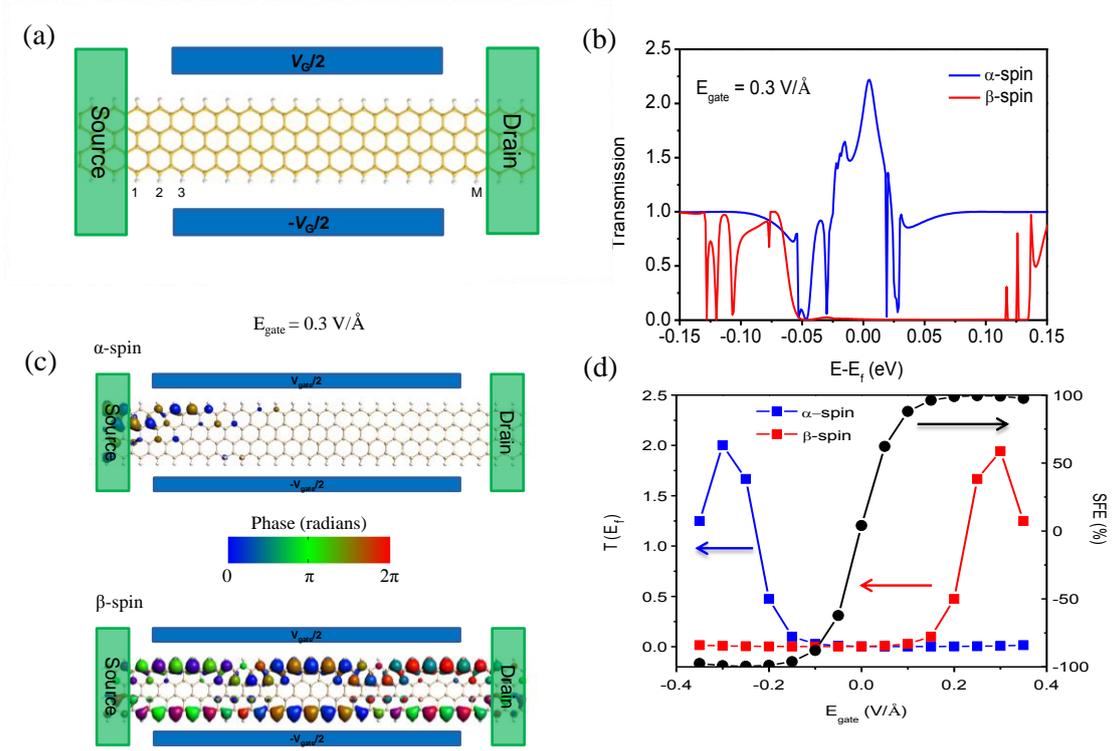

Fig. 3 Spin-filter based on the 4-ZSiNR. (a) Schematic model with one pair of gate electrodes on the two sides. (b) Spin-resolved transport spectrum under $E_{gate}$ = 0.3 V/Å. (c) Spin-resolved transmission eigenstate at $E_f$ and the Γ point in $k$-space under $E_{gate}$ = 0.3 V/Å. The isovalue is 1.0 a.u. (d) Spin-resolved transmission coefficient at $E_f$ $k$-space and spin filtration efficiency as a function of $E_{gate}$.

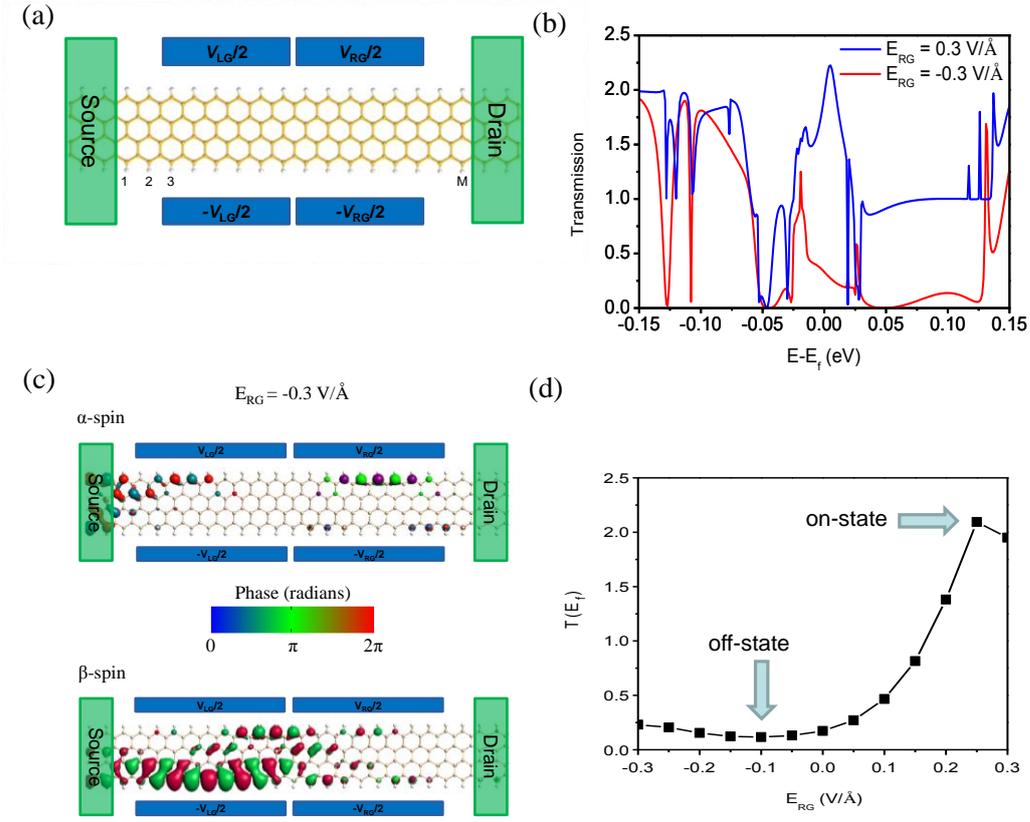

Fig. 4 Spin field effect transistor based on the 4-ZSiNR. (a) Schematic model with two pairs of gate electrodes on the two sides. The electrical field of the left pair of electrodes is fixed (-0.3 V/Å), while the right pair of electrodes switches the current on and off. (b) Total transport spectrum under $E_{RG} = \pm 0.3$ V/Å. (c) Spin-resolved transmission eigenstate at $E_f$ and the Γ point in $k$-space under $E_{RG}$ = -0.3 V/Å and that under $E_{RG}$ = -0.3 V/Å can be found in Fig. 3(c). The isovalue is 1.0 a.u. (d) Total transmission coefficient at $E_f$ as a function of $E_{RG}$.